\documentclass[12pt]{article}

\pdfoutput=1
\usepackage{subfigure}
\usepackage{amssymb,amsmath}
\usepackage{graphicx}
\usepackage{color}
\usepackage[colorlinks=true
,urlcolor=blue
,citecolor=blue
,linkcolor=blue
,pagecolor=blue
,linktocpage=true
,pdfproducer=medialab
]{hyperref}
\usepackage[a4paper,width=15.5cm]{geometry}
\makeatletter \renewcommand{\@dotsep}{10000} \makeatother
\usepackage{appendix}

\usepackage{cite}

\newcommand{\beq}{\begin{equation}}
\newcommand{\eeq}{\end{equation}}
\newcommand{\bea}{\begin{eqnarray}}
\newcommand{\eea}{\end{eqnarray}}

\def\lrprtmu{\stackrel{\leftrightarrow}{\partial_\mu}}
\def\half{{\textstyle{1\over 2}}}

\begin{document}

\begin{center}

 {\Large\bf  Lorentz violation and Condensed Matter Systems
 } \vspace{1cm}

{\large   Muhammad Adeel Ajaib\footnote{ E-mail: adeel@udel.edu}}

{\baselineskip 20pt \it
University of Delaware, Newark, DE 19716, USA  \\Department of Mathematics, Statistics and Physics, Qatar  University,  Doha,  Qatar
} \vspace{.5cm}

\vspace{1.5cm}
\end{center}

\begin{abstract}

We present heuristic arguments that hint to a possible connection of Lorentz violation with observed phenomenon in condensed matter physics.
Various references from condensed matter literature are cited where operators in the Standard Model Extension appear to be enhanced.
Furthermore, we consider the L{\'e}vy-Leblond equation, which is the analogue of Dirac equation in non-relativistic quantum mechanics. We show that we can obtain the L{\'e}vy-Leblond equation by adding enhanced Lorentz violating operators to the Dirac equation. Based on these observations, we propose that Lorentz violation exhibits itself in non-relativistic quantum mechanics.

\end{abstract}

\newpage

\section{Introduction}

Einstein's theory of relativity has stood the test of time and is in excellent agreement with observations. Various attempts have been made to develop scenarios where Lorentz violation can exhibit itself in nature \cite{Colladay, Pavlopoulos:1967dm, Mattingly:2005re}. The Standard Model Extension (SME) \cite{Colladay} provides a consistent framework for incorporating Lorentz violation as a low energy effective theory. The effective operators in the SME involve couplings of fermions with background fields which transform under observer Lorentz transformation but remain invariant under particle Lorentz transformations \cite{Colladay}.

This article aims to highlight an interesting similarity between developments in condensed matter physics and Lorentz violation in the last few decades. In particular, we focus on Lorentz violation as understood in the SME. In reference \cite{Ajaib:2012wq} it was pointed out that, in the non-relativistic limit, one of the terms in the SME yields the Rashba spin-orbit interaction that arise in condensed matter systems with structural inversion asymmetry. In this article we point out several terms in the SME that are employed in developing fields in condensed matter physics, such as spintronics and the study of graphene. Furthermore, we consider the L{\'e}vy-Leblond Equation  (LLE) \cite{LevyLeblond:1967zz, Ajaib:2015uha,Ajaib:2015eer, Ajaib:2017cpl} and show that it can be obtained by adding Lorentz violating terms in the Dirac equation. Based on these observations, we propose that Lorentz violation exhibits itself in the non-relativistic limit.

The paper is organized as follows: In section \ref{sme} we briefly introduce the SME. In section \ref{hamiltonian} we present the non-relativistic Hamiltonian of the SME. We discuss the terms in the SME that are typically considered in various condensed matter systems. In section \ref{sec:lle} we consider the LLE and show that it also be obtained by adding Lorentz violating terms in the Dirac equation. 
Section \ref{conclusion} outlines our conclusion.

\section{Brief review of the SME}\label{sme}

The minimal SME provides a standard framework for testing Lorentz and CPT violation in various experiments \cite{Kostelecky:2008ts} (for reviews see \cite{Bluhm:2005uj}). The QED sector of the mSME Lagrangian for a single species of fermion with mass $m$ is given by ($\hbar=c=1$),
\begin{eqnarray}
{\cal L} =  {i \over 2} \bar{\psi}\Gamma^\mu \lrprtmu \psi -
\bar{\psi}M \psi ,
\label{smelag}
\end{eqnarray}
where, $A \lrprtmu B \equiv A \partial_\mu B - (\partial_\mu A) B $ and $\mu=0,1,2,3$.
\begin{eqnarray}
\Gamma^\nu & = & \gamma^\nu + c^{\mu\nu}\gamma_\mu + d^{\mu\nu}\gamma_5 \gamma_\mu
+ e^\nu + if^{\nu}\gamma_5 + \frac12 g^{\lambda \mu \nu}
\sigma_{\lambda \mu}, \label{Gamma} \\
M & = & m + a_\mu\gamma^\mu + b_\mu\gamma_5 \gamma^\mu + \frac12
H_{\mu\nu}\sigma^{\mu\nu}\label{Mass}.
\end{eqnarray}
The coefficients of the SME contained in $\Gamma^\nu$ are dimensionless whereas those in $M$ have dimensions of mass. The terms with the coefficients $a_{\mu}$, $b_{\mu}$, $e_{\mu}$, $f_{\mu}$ and $g_{\lambda \mu \nu}$ violate CPT while those with coefficients $c_{\mu \nu}$, $d_{\mu \nu}$ and $H_{\mu \nu}$ preserve CPT.
These background fields transform in their respective manners (e.g. as vectors, tensors, etc.) under observer Lorentz transformations so that the  Lagrangian is Lorentz invariant under observer Lorentz transformations. Under particle Lorentz transformations only the fermion fields are boosted or rotated and the background fields given in equations (\ref{Gamma}) and (\ref{Mass}) remain invariant. Therefore, the terms in the mSME lead to particle Lorentz violation. 
The formalism of the SME allows for observer transformations that enables the choice of any combination of these coefficients. We will see that various condensed matter systems corresponds to enhanced values of different combinations of these coefficients.

\begin{table*}[t]
\begin{tabular}{|p{3.06cm}|c|p{1.25cm}|c|} \hline \hline
System & $H_{CM}$ &  {$h_{NR}$ {\footnotesize coeff.}}
& References   \\ \hline \hline
Rashba SOI &
$ \begin{array}{c}
\frac{\alpha}{\hbar}\, \left(p_x \sigma_y - p_y \sigma_x \right)
\end{array}$    &
$D_{jk}$   &
\cite{Rashba:1960, Rashba:1984, spintronics-review:cikkek,review_spinHall:cikkek}
\\[3mm]  \hline 
Dresselhaus SOI &
$ \begin{array}{c}
 \frac{\beta}{\hbar}\, \left(p_y \sigma_y - p_x \sigma_x \right)
\end{array}$    &
$D_{jk}$   &
\cite{Dresselhaus:1955, spintronics-review:cikkek,review_spinHall:cikkek}
\\[3mm] \hline
Single-layer 
graphene
&
$v \left(p_x \sigma_x + p_y \sigma_y \right)$ &
$D_{jk}$  &
\cite{McClure_DiVincenzo:cikk,Novoselov-1:cikk,Kim:cikk,Graphene:cikkek,Katsnelson:cikk}  \\[3mm] \hline
Bilayer graphene &
$ \frac{1}{2m} \left(\frac{p_+^2 +p_-^2 }{2}\sigma_x
-\frac{p_-^2 - p_+^2 }{2i}\sigma_y\right)$ &
$F_{jkl}$   &
\cite{Novoselov-2:cikk,Ed-Falko:cikk,Katsnelson:cikk} \\ \hline

Heavy holes 
in a quantum well
&
$  i \frac{\tilde{\alpha}}{2\hbar^3}
\left(p_-^3 \sigma_+ - p_+^3 \sigma_- \right)$    &
$I_{jklm}$  &
\cite{Winkler_Schliemann-heavy-hole:cikkek,review_spinHall:cikkek}   \\[3mm] \hline
Bulk Dresselhaus &
$
\begin{array}{c}
\frac{\gamma_\textrm{D}}{\hbar^3}\left[
\sigma_x p_x\left(p_y^2 - p_z^2 \right)
+\sigma_y p_y\left(p_z^2 - p_x^2 \right) \right.
\\[1ex]
\left.
+\sigma_z p_z\left(p_x^2 - p_y^2 \right)
\right]
\end{array}
$ &
$I_{jklm}$  &
\cite{spintronics-review:cikkek,review_spinHall:cikkek}  
\\
 \hline \hline
\end{tabular}
\caption{ We list here some common interaction Hamiltonians employed in condensed matter literature and the relevant coefficients of the SME. The coefficients shown in the third column are linear combinations of the SME coefficients displayed in equations (\ref{A})-(\ref{I}). Here $p_\pm = p_x \pm i p_y$, $\sigma_\pm = \sigma_x  \pm i \sigma_y$. This table has been adopted from reference \cite{Cserti:2006} in order to highlight the relevant coefficients in the SME. For the complete table the reader is referred to this reference. 
\label{table}}
\end{table*}

\section{The Hamiltonian}\label{hamiltonian}

In this section we present the non-relativistic Hamiltonian of the SME derived in ref \cite{Kostelecky:1999zh} and discuss various examples in condensed matter where these terms are enhanced. The systems we discuss will not be unique and there can be other systems where similar interactions might be realized. These examples will however render basis to the proposition that Lorentz violation is manifested in various condensed matter systems.

Table \ref{table} shows various examples of Hamiltonians employed in condensed matter systems. The reader is referred to the references provided in the table for further details. Below we present the non-relativistic Hamiltonian of the SME derived in  \cite{Kostelecky:1999zh} using the Foldy-Wouthuysen transformation and discuss the terms of the SME relevant for various condensed matter systems. The non-relativistic Hamiltonian $h_{NR}$ for the two component fermion is given by \cite{Kostelecky:1999zh, Lehnert:2004ri}
\begin{eqnarray}
h_{NR}&=& m+\frac{p^2}{2m}+A+B_j \sigma_j +C_j \ \frac{p_j}{m}+D_{jk} \ \frac{p_j \sigma_k}{m} +E_{jk} \ \frac{p_j p_k}{m^2} \nonumber \\[2mm]
&+& F_{jkl}\ \frac{p_j p_k \sigma_l}{m^2} +G_{jkl}\ \frac{p_j p_k p_l}{m^3}+I_{jklm} \ \frac{p_j p_k p_l \sigma_m}{m^3}
\label{hnr}
\end{eqnarray}
where
\begin{eqnarray}
A&=& a_0-mc_{00}-m e_0  \label{A} \\[2mm]
B_j &=& -b_j+md_{j0} -\frac{1}{2} m \epsilon_{jkl} g_{kl0}+\frac{1}{2} \epsilon_{jkl} H_{kl}  \label{B} \\[2mm]
C_j &=& -a_j+m(c_{0j}+c_{j0})+m e_j  \\[2mm]
D_{jk} &=& b_0 \delta_{j k} - m(d_{jk}+d_{00} \delta_{jk})- m \epsilon_{klm}(\half g_{mlj}+g_{m00}\delta_{jl}) -\epsilon_{jkl} H_{l0}  \label{D} \\[2mm]
E_{jk} &=& m(-c_{jk}-\frac{1}{2} c_{00}\delta_{jk}) \\[2mm]
F_{jkl} &=& [ m(d_{0j}+d_{j0})
       -\half( b_j+md_{j0}+\half m\epsilon_{jmn}g_{mn0}
       +\half \epsilon_{jmn}H_{mn} ) ]\delta_{kl} 
       \nonumber \\
       & &+\half \left(b_l+\half m\epsilon_{lmn}g_{mn0}\right)\delta_{jk}
       -m \epsilon_{jlm} (g_{m0k}+g_{mk0})    \\[2mm]
G_{jkl} &=& a_j\delta_{kl}-me_j\delta_{kl}  \\[2mm]
I_{jklm} &=& \half\left[
   \left( -b_0\delta_{jm}+m d_{mj}+\epsilon_{jmn}H_{n0}
    \right) \delta_{kl}
      +\left( -m d_{jk}-\half m\delta_{knp} g_{npj}
   \right) \delta_{lm}
   \right] \label{I} 
\end{eqnarray}
Here $\sigma_i$ denote the three Pauli matrices. The implications of various coefficients in the non-relativistic limit has been studied extensively in the SME (see, for example \cite{Bluhm:2000gv} and reference \cite{Kostelecky:2008ts} for limits on various coefficients of the SME).  Here we discuss the condensed matter systems that exhibit enhanced values of these coefficients. Following are some of the important interactions and materials employed in contemporary condensed matter physics relevant to the SME framework:

\begin{itemize}

\item \textbf{Rashba interaction \cite{Rashba:1960, Rashba:1984}:} Spin orbit interaction arises in the non-relativistic limit of the Dirac equation. These interactions are particularly important in spintronics and can influence the properties of the electrons in these systems. An important type of SOI, called the Rashba SOI arises due to structural inversion asymmetry  at the interface of, for example, semiconductor hetrostructures. Due to the asymmetry in the confining potential an electric field is induced at the interface. A 2 dimensional electron gas (2DEG) at the interface `sees' this electric field as an effective magnetic field and this lifts the spin degeneracy of the electrons at the interface.
The Rashba Hamiltonian for a 2DEG confined in the $xy$ plane is given by
\begin{eqnarray}
H_R=\frac{\alpha_R}{\hbar} (\sigma_x p_y -\sigma_y p_x)
\label{rashba}
\end{eqnarray}
where $\alpha_R$ is the Rashba coupling and its experimental value of the order $10^{-11}-10^{-10}$ eV-m. The coefficient $\alpha_R$ is proportional to the confining electric field $E_z$ which quantifies the asymmetry in the confining potential. 

In the SME this coefficient is proportional to the relevant components of the background fields ($d$, $g$ and $H$) contained in the coefficient $D_{jk}$ in equation (\ref{D}).
It was shown in reference  \cite{Ajaib:2012wq}	that a Rashba-type interaction term also arises in the non-relativistic limit of the SME. 
 A limit was thereby derived for Rashba coefficient for Lorentz violation ($\alpha_{RLV} =\hbar c H_{03}/m \lesssim 10^{-30}$ eV-m). Similarly, we can also derive the SME coefficient for condensed matter systems. If we choose the typical value of $\alpha_R \sim 10^{-10}$ eV-m for a 2 dimensional electron gas with interaction Hamiltonian in equation (\ref{rashba}), the SME coefficient $D_{12} \sim 10^{-6}$ GeV, which is around 20 orders of magnitude larger than the current limit for, say, $D_{12}=H_{03}$.

\item \textbf{Dresselhaus interaction \cite{Dresselhaus:1955, spintronics-review:cikkek,review_spinHall:cikkek} :} This type of spin-orbit interaction also plays an important role in spintronics. It is understood to arise as a result of the bulk inversion asymmetry in the material. The Hamiltonian for cubic Dresselhaus interaction is given by
\begin{eqnarray}
H_{D}^{(3)}=\frac{\gamma_D}{\hbar^3}[\sigma_x p_x (p_y^2-p_z^2)
+\sigma_y p_y (p_z^2-p_x^2)
+\sigma_z p_z (p_x^2-p_y^2)
]
\end{eqnarray}
where $\gamma_D$ is the Dresselhaus constant and its value is material dependent. Experimentally the value of $\gamma_D \sim 10^{-30} \ {\rm eV  m^3}$. The corresponding SME coefficient for cubic Dresselhaus interaction is $I_{jklm}$. The relationship of $\gamma_D$ with, say, $I_{jjkk}=m d_{jj}$ is
\begin{eqnarray}
\gamma_D = \frac{\hbar^3 d_{jj}}{m^2 c}
\end{eqnarray}
Therefore, for $\gamma_D \sim 10^{-30} \ {\rm eV  m^3}$ we get $d_{jj} \sim 10^2$ for the electron. In a narrow quantum well, the operators $p_z$ and $p_z^2$ can be replaced by their expectation values.  This yields the linear Dresselhaus interaction as 
\begin{eqnarray}
H_D=\beta (\sigma_x p_x -\sigma_y p_y)
\end{eqnarray}
where $\beta=\gamma_D \langle p_z^2\rangle$. The cubic part of the above Hamiltonian can be neglected for strong confinement along the $z$ direction. The typical value of $\beta \sim 10^{-11}$ eV-m.

 It is interesting to note that Rashba and Dresselhaus interactions depend on the asymmetry of the structure. Absence of symmetry in condensed matter systems typically leads to such interactions as a result of induced effective (electric or magnetic) fields.

\item \textbf{Graphene \cite{McClure_DiVincenzo:cikk,Novoselov-1:cikk,Kim:cikk,Graphene:cikkek,Katsnelson:cikk} :} Graphene is typically referred to a material that is essentially 2 dimensional and is composed of a single layer of carbon atoms arranged in a honey comb lattice. It is one of the most important materials in contemporary condensed matter physics due to the very interesting properties it possesses. These properties include high electron mobility, thermal conductivity and a high damage threshold. Another interesting property is that its opacity solely depends on the fine structure constant \cite{Nair:2008zz}.

In graphene, quasi-particles near the Dirac point behave as 2 dimensional massless Dirac fermions. The dispersion relation of these particles is linear, $E=\pm v_F p$. The Hamiltonian of these particles is that of a massless spin 1/2 particle given by
\begin{eqnarray}
H_{G}=v_F \vec{\sigma}.\vec{p}
\label{hg}
\end{eqnarray}
where $v_F \sim 10^{6} \ m/s$ is the Fermi velocity and $\vec{p}$ is the momentum of the quasi-particle. The above Hamiltonian has the form of Weyl equation which arises in the massless limit of the Dirac equation. This Hamiltonian is a consequence of graphene's crystal symmetry. 

The SME framework can also be used to explain the origin of the Hamiltonian in equation (\ref{hg}). The coefficient $c_{00}$ in the SME is unique since the term with this coefficient has the form of the kinetic energy of the particle. If the only non-zero coefficients in the SME are $c_{00} (=1)$ and $d_{00}$ then the Hamiltonian in equation (\ref{hnr}) takes the form
\begin{eqnarray}
h_{NR}=(d_{00}c) \ \vec{\sigma}.\vec{p}
\end{eqnarray}
where $c$ is the speed of light. Therefore 
\begin{eqnarray}
v_F=d_{00}\ c
\end{eqnarray}
An enhancement in the coefficient $c_{00}$ and $|d_{00}| \sim 10^{-2}$ might account for the behavior of quasi-particles in graphene. Similarly, for the Hamiltonian of bilayer graphene, displayed in the fourth row of Table \ref{table}, additional contribution to the Hamiltonian arise from the coefficient $F_{jkl}$ of the SME.

Stability and causality issues for relativistic field theories with massive fermions and a non-zero $c_{\mu \nu}$ coefficient was studied in ref \cite{Kostelecky:2000mm}. It was shown that a non-zero $c_{\mu \nu}$ can lead to issues like instability and causality violations in the theory. In particular a positive $c_{00}$ implies issues with stability (space-like momenta) while a negative $c_{00}$ can lead to microcausality violations. 
This however can be avoided by appropriate modification of the dispersion relations \cite{Kostelecky:2000mm}. For the case of graphene these conclusions do not necessarily hold since the fermions are essentially massless and the momenta involved are non-relativistic. Note also that a non-zero $c_{\mu \nu}$ is equivalent to a model with non-zero $\tilde{k}_{\mu \nu}$ (see, for example references \cite{Hohensee:2008xz} and \cite{Hohensee:2009zk} for details).

\item \textbf{Heavy holes:} Another example is a system with heavy holes where a structure inversion asymmetry similar to the Rashba interaction leads to the cubic  interaction displayed in fifth row Table \ref{table}. This again is accounted for by the coefficient $I_{jklm}$ provided in equation (\ref{I}). The reader is referred to the references in the table for further details on the materials where these interactions are important.

\end{itemize}

 The coefficients in the SME are understood as background fields that couple to the fermions. 
In condensed matter systems the role of these background fields are played by physical parameters such as the electric and magnetic fields. It is also interesting that in most cases the important SME terms in condensed matter systems are connected to the symmetry of the structure.

The above examples can serve as motivation to further investigate systems where the dynamics of the carriers are described by enhanced SME coefficients and other linear combinations. 
The ideas presented herein can be tested in several ways. If this proposal is correct than it should be possible to realize situations in condensed matter systems where other coefficients of the SME get enhanced. Therefore future investigation is required to test this proposition. 

\section{Lorentz violation and the L{\'e}vy-Leblond Equation}\label{sec:lle}

The LLE is the analogue of the Dirac equation and describes fermions in the non-relativistic limit \cite{LevyLeblond:1967zz, Ajaib:2015uha,Ajaib:2015eer, Ajaib:2017cpl}. In this section we discuss how the LLE can be viewed from the perspective of Lorentz violation. It was shown in \cite{Ajaib:2015eer} that the LLE is the non-relativistic limit of the Dirac equation. The Dirac equation is Lorentz invariant whereas the LLE is invariant under Galilean transformations \cite{LevyLeblond:1967zz}.  We show that we can obtain the LLE by adding operators, that violate Lorentz invariance, to the Dirac equation. 
The Dirac equation is given by
\begin{eqnarray}
(i   \gamma^\mu \partial_\mu  -  m) \psi =0
\label{DE}
\end{eqnarray}
or
\begin{eqnarray}
(i   \gamma_0 \partial_0+i \gamma_i \partial_i   -  m) \psi =0
\end{eqnarray}
where $\gamma_\mu$ are the Dirac matrices. Considering plane wave solutions  ($\psi=u(p) e^{-i p.x}=u(p) e^{-i (E t - p_z z )}$) and following the procedure outlined in reference \cite{Ajaib:2015eer} we obtain the non-relativistic limit of the Dirac equation as
\begin{eqnarray}
\left( \eta_1 E' - \gamma_i p_i + \eta_2 m \right) u =0
\label{eq:nrde}
\end{eqnarray}
Here $\eta_{1}=(\gamma_0+ I)/2$ and $\eta_{2}=(\gamma_0 - I)$ and the above equation yields the dispersion relation of a non-relativistic particle ($E'=p_i^2/2m$). The matrices $\eta_{1,2}$ are hermitian and are singular ($\eta_1\eta_2=0$). Note that in \cite{Ajaib:2015eer} the non-relativistic limit of the Dirac equation was analyzed with the $i\gamma_5$ mass term. The matrices obtained in \cite{Ajaib:2015eer} were non-hermitian and singular. 
Equation (\ref{eq:nrde}) can be written as
\begin{eqnarray}
\left( i \eta^\mu \partial_\mu + \eta_2 m \right) \psi =0 
\label{aeq}
\end{eqnarray}
where  $\eta^\mu=(\eta^0,\eta^i)=(\eta_1,\gamma^i)$.
Equation (\ref{eq:nrde}) was obtained as the non-relativistic limit of the Dirac equation. We can also view equation (\ref{eq:nrde}) from the perspective of Lorentz violation. 
For that, consider the Dirac Lagrangian (${\cal L}_{D}$) and the  Lagrangian corresponding to the LLE (${\cal L}_{L}$),
\begin{eqnarray}
{\cal L}_{D}=i \bar{\psi} \gamma^\mu \partial_\mu \psi - m  \bar{\psi}\psi
\end{eqnarray}
\begin{eqnarray}
{\cal L}_{L}=i \bar{\psi} \eta^\mu \partial_\mu \psi + m \bar{\psi} \eta_2 \psi
\end{eqnarray}
The Lagrangian corresponding to the LLE can be written as 
\begin{eqnarray}
{\cal L}_{L}={\cal L}_{D}+{\cal L}_{LV}
\end{eqnarray}
where ${\cal L}_{LV}$ includes terms that violate Lorentz invariance. These terms, when added to the Dirac Lagrangian, yield the Lagrangian corresponding to (\ref{eq:nrde}), i.e.,
\begin{eqnarray}
{\cal L}_{LV}=i \bar{\psi} \left( \frac{I-\gamma_0}{2} \right) \partial_0 \psi + m \bar{\psi} \gamma_0 \psi
\end{eqnarray}
We can see that the Lorentz violating terms have enhanced coefficients. This further supports the observations made in the previous section, i.e., enhanced Lorentz violating operators play an important role in non-relativistic quantum mechanics.

\section{Conclusion}\label{conclusion}

We suggested that several effects observed in condensed matter systems can be attributed to enhancement of Lorentz violation in the SME. 
Various references were presented as examples where these coefficients are enhanced.  Further analysis is required to test this proposition. 
The coefficients of the SME are realized as various physical parameters in condensed matter systems. For example, systems where Rashba and Dresselhauss interactions are realized are examples where the coefficients of the SME play the role of fields induced due to the asymmetry of the structure (spatial inversion  or bulk inversion asymmetry). Similarly, graphene is an example of a material where the crystal symmetry leads to a Hamiltonian for the quasi-particles that can be obtained in the SME. 

We also considered the L{\'e}vy-Leblond equation and showed that it can be obtained in the non-relativistic limit of the Dirac equation. In addition, we also showed that the L{\'e}vy-Leblond equation can be obtained by adding Lorentz violating operators to the Dirac equation. Based on these observations, we propose that Lorentz violation exhibits itself in non-relativistic quantum mechanics.

\section{Acknowledgments}

The author would like to thank Alan Kostelecky, Fariha Nasir, Asif Warsi and Yang Zhou for useful discussions.

\end{document}